# Antibacterial size effect of ZnO nanoparticles and their role as additives in emulsion waterborne paint

Imroi El-Habib[a,b,*], Hassan Maatouk[b], Alex Lemarchand[a], Anne Roynette[b], Sarah Dine[a], Christine Mielcarek[b], Mamadou Traore[a], Rabah Azouani[b,*]

[a] Laboratoire des Sciences des Procédés et des Matériaux (LSPM-CNRS UPR 3407), Institut Galilée, Université Sorbonne Paris Nord, 99 Avenue Jean-Baptiste Clément, 93430 Villetaneuse, France

[b] Ecole de Biologie Industrielle (EBI), 49 Avenue des Genottes – CS 90009, F95895 Cergy Cedex, France

ABSTRACT

Nosocomial infections (NIs) are prevalent in intensive care units due to antibiotic overuse. Metal oxide nanoparticles (NPs), like ZnO, offer potential solutions, yet understanding how NPs size impacts their antibacterial efficacy are lacking. This study focuses on the effect of nanoparticle size on kinetics of bacterial strains growth. NPs were synthesized using a sol-gel process with monoethanolamine (MEA) and water, characterized using X-ray diffraction (XRD), transmission electron microscopy (TEM), and Raman spectroscopy, confirming crystallization and size variations. ZnO NPs with mean size of 22, 35 and 66 nm were used against the most common nosocomial bacteria strains *Escherichia coli* (Gram-negative), *Pseudomonas aeruginosa* (Gram-negative), and *Staphylococcus aureus* (Gram-positive). The evaluation of NPs minimal inhibitory concentration (MIC) and bactericidal concentration (MBC) revealed superior antibacterial activity in smaller NPs. The bacterial population was monitored via optical absorbance, showing reduced specific growth rate, prolonged latency period, and increased inhibition percentage with smaller NPs, indicating a substantial deceleration in the growth of microorganisms. *Pseudomonas aeruginosa* exhibited the smallest sensitivity to ZnO NPs, attributed to its environmental stress resistance. Furthermore, the antibacterial efficacy of paint containing 1 wt% of 22 nm ZnO NPs was assessed and displayed activity against *E. coli* and *S. aureus*.

[*] Corresponding authors.
E-mail adresses : el-habib.imroi@lspm.cnrs.fr (I. El-Habib), r.azouani@hubebi.com (R. Azouani)

1. **Introduction**

Nosocomial infections (NIs) represent common complications among patients admitted to intensive care units (ICUs), with reported incidences ranging from 5% to 10% in Europe and America [1]. These infections result from antibiotic overuse, leading to microbial resistance. Consequently, considerable attention within the pharmaceutical and medical sectors has been directed towards developing new preventive strategies. This includes exploring alternative methodologies beyond the development of novel antibiotics and the implementation of regulated bio-decontamination procedures, all aimed at significantly reducing the incidence of infections.

Metal oxide nanoparticles (NPs) have emerged as promising candidates in this endeavor, offering a potential solution to prevent the emergence of resistant pathogens [2]. The inherent physicochemical properties of metal oxide NPs, coupled with their high surface-to-volume ratio, enable them to exert diverse modes of action (MOAs) against microorganisms. These mechanisms include the release of reactive oxygen species (ROS), the liberation of ions, and direct interaction with bacterial cell walls [3]. Consequently, metallic oxide nanoparticles hold promise as potent antimicrobial agents [4,5], owing to their ability to use multiple mechanisms of action in combating bacterial resistance.

Zinc oxide garners attention as a compelling metal oxide material due to its biocompatibility, easy synthesis, chemical stability, high abundancy and affordability [6–8]. These characteristics make it ideal candidate for various biomedical applications, including its potential use in combating bacterial infections. Previous studies have demonstrated the effectiveness of ZnO nanoparticles against a broad spectrum of pathogenic microorganisms, including both Gram-positive and Gram-negative bacteria [9–11]. Their mode of action relies, in part, on the production of reactive oxygen species [12–16], initiating the oxidation of cellular components, disruption of bacterial membranes [17–22], direct contact [18,23–25] and internalization [21, 26, 27], ultimately culminating in cell death.

Enhancing the antibacterial potential of ZnO NPs is crucial for effectively preventing infections. The existing literature provides insights into how ZnO nanoparticle characteristics, such as shape [28–33], size [15,34–37], concentration [38,39] and surface modification [40–43], impact their antimicrobial activity. However, our study focus on the effect of the size nanoparticle on the kinetics of bacterial strains growth. Indeed, a smaller size implies a high surface area of the materials. This increase in specific surface area makes them more reactive, resulting in a substantial concentration of surface particles capable of interacting with electrons and holes. This unique characteristic offers more sites where electric charges can accumulate, resulting in an increased surface electric charge, which leads to strong electrostatic interactions between ZnO NPs and bacterial membranes. The particle's charge can influence its biocompatibility and ability to traverse biological barriers [44]. The small dimension of materials also makes them more abrasive [45]. Additionally, the sizes of NPs (ranging from 1 to 100 nm) are comparable to the size of protein globules (2 to 10 nm), the diameter of the DNA helix (2 nm), and the thickness of cell membranes (10 nm) [46]. Thus, the decrease in their size allows them to easily enter cells and cell organelles.

Previous studies have demonstrated that reducing nanoparticle size enhances their antibacterial activity [15,34–37]. Babayevska et al. [15] found that ZnO nanoparticles showed greater effectiveness compared to microparticles against *Escherichia coli* and *Staphylococcus aureus*. They explained this difference by examining the mechanism through ROS measurements and found that NPs produced more ROS than microparticles. Nair et al. [47] demonstrated that the antibacterial activity of ZnO NPs (ranging from 40 nm to 1.2 μm) is higher when their size is smaller. They analyzed the mode of action of only one ZnO nanoparticles size by conducting scanning electron microscope (SEM) analysis on both treated and untreated *E. coli* cells and found bacteria cell destruction. Raghupathi et al. [35] have discovered a superior antibacterial activity of ZnO NPs at reduced sizes. They compared the growth curves of bacteria in the presence of ZnO NPs of various sizes (30 nm, 88 nm, 142 nm, and 212 nm) at a concentration of 6 mM during 6-hour cultures. Despite of these efforts, there is limited research comparing the bacterial growth kinetic across different sizes to understand the size-dependent mode of action.

To gain new insights into the role of NPs size on their antibacterial activity, this study aims to investigate the impact of ZnO NPs size on the growth kinetic of both Gram-negative and Gram-positive bacteria during a 24-hour culture period in the presence of a wide range of NPs concentrations. The bacterial growth kinetic will be examined in the presence of various NPs sizes by monitoring the change in bacterial optical density at different NPs concentrations and determining the specific growth rate of bacteria, their latency period, along with the percentage inhibition of NPs. Prior to this, we will assess the impact of the size of our synthesized ZnO NPs on their antibacterial activity by determining their minimal inhibitory concentration (MIC) and bactericidal concentration (MBC). Finally, we will evaluate the conservation of the antibacterial activity of synthesized ZnO NPs in a paint formulation as biocidal agent.

## 2. Material and Methods

### 2.1. Material

Zinc acetate dihydrate (≥ 99%), butan-1-ol (≥ 99.5%), acetone (≥ 99.5%), and silicone oil were obtained from Sigma-Aldrich. Isopropanol (≥ 99.5%) was acquired from Acros Organics. Monoethanolamine (100%) was sourced from Emprove. Sodium polyacrylate (PAAS) was supplied by Cosmedia sp. Muller Hinton broth and Trypto Soybean Casein was purchased from DIFCOTM and BioMerieux, respectively. Powder paint was sourced from Dolci, and Eugon LT SUP was acquired from BioMerieux.

### 2.2. Synthesis of ZnO nanoparticles

The protocol applied to obtain ZnO nanopowders of various sizes was described in detail in our recent article [48]. A zinc precursor, with a concentration of 0.1 M, was dissolved in 100 mL butan-1-ol in the presence of a complexing agent, monoethanolamine (MEA). Water was introduced into the reaction medium to induce NPs precipitation and obtain nanopowders with satisfying yield. NPs size was mainly controlled by the ratio of zinc ions to complexing agent $[Zn^{2+}]/[MEA]$ and the hydrolysis rate $[Zn^{2+}]/[H_2O]$. Nanopowders were collected by centrifugation, followed by three washes with isopropanol and acetone. Next, 2 wt% nanopowders were dispersed in Mueller-Hinton broth using ultrasonication, with 0.4 wt% PAAS serving as a dispersant.

The synthesis parameters used to prepare ZnO nanopowders are summarized in Table 1.

Table 1 : Synthesis parameters of ZnO nanopowders.

| Sample size | $[Zn^{2+}]$/ [MEA] | $[Zn^{2+}]$/[H2O] | Agitation time | Temperature | Solvent |
|---|---|---|---|---|---|
| 22 nm | 1 | 5 | 22 h | 80 °C | |
| 35 nm | 1 | 10 | 2 h | 110°C | Butan-1-ol |
| 66 nm | 2 | 10 | 22 h | 110°C | |

### 2.3. Characterization of ZnO nanoparticles

Phase identification and structural and microstructural characterization of ZnO nanopowders were performed by X-ray diffraction (XRD) measurements on an INEL XRG 3000 diffractometer using a monochromatic cobalt source (λ Kα1 (Co) = 1.788976 Å). Diffraction patterns were processed using the Rietveld method implemented in MAUD (Material Analysis Using Diffraction) software. Pattern was fitted according to a standardized procedure. The parameters refined were cell parameters, NPs size and the presence of micro deformation. The crystallographic information file for ZnO wurtzite was obtained from Crystallography Open Database (Ref 2300450). In a final step, in order to optimize the quality of the Rietveld pattern refinement, an arbitrary texture option was used. This characterization was completed by Transmission Electronic Microscopy performed on a JEOL 2011 equipped with a Gatan Imaging Filter (DIF) 200. Raman spectroscopy measurements were also carried out on the powder samples using a HORIBA Jobin-Yvon HR800 spectrometer with an excitation wavelength (λ = 633 nm) to confirm the formation of ZnO NPs.

### 2.4. Minimal Inhibitory Concentration (MIC)

The produced ZnO nanopowders were dispersed in Muller Hinton broth in the presence of a dispersant, the PAAS, for subsequent antibacterial activity studies. NPs solutions underwent ultrasonication in J.P. Selecta ultrasonic bath at 150 W and 42 kHz for 1h. The protocol for preparing stable NPs suspensions was inspired by that of Luo et al. [49].

Subsequently, microdilutions of the ZnO NPs suspension were performed, without cascade, to obtain different concentrations of NPs. This dilution took place in a dispersant solution composed of PAAS and culture medium to maintain the concentration of the dispersant. The samples were then inoculated on a 96-well microplate in contact with microorganisms. The total volume of a well was 210 µL including 200 µL of NPs suspensions and 10 µL of microorganism suspension. The inoculated concentration of all tested microorganisms was 6 $\mathrm{Log}_{10}$ UFC/mL. It was prepared from optical density calibration. The negative controls were the culture medium, NPs solution and sodium polyacrylate solution. The positive control were the dispersant solution in contact with microorganisms without NPs and the microorganisms in the culture medium alone without dispersant either NPs. The experiences were performed in triplicate.

Bacterial strains examined included *E. coli* (ATCC 8739), *P. aeruginosa* (ATCC 9027), and *S. aureus* (ATCC 6538). Incubation temperatures were set at 37°C for *E. coli* and *S. aureus*, and at 30°C for *P. aeruginosa*. Incubation time for all bacteria was 24 hours. The 96-well microplate was incubated in a FLUOstar Omega spectrophotometer. Microorganisms were cultured from cryotubes stored at -80°C.

Additionally, physiological water was added to empty wells during microplate preparation to ensure proper humidity.

The MIC corresponds to the lowest concentration of NPs that prevents visible growth of bacteria after the incubation period. It was determined visually and confirmed by the evolution of optical density at 600 nm ($OD_{600}$) on the FLUOstar Omega spectrophotometer.

### 2.5. Minimal Bactericidal Concentration (MBC)

The MBC is the lowest concentration of an antibacterial agent that results in at least a 99.9% decrease in cell viability. After incubation in FLUOstar, the samples in the microplate wells were plated on Muller Hinton (MH) agar and incubated at 37°C for *E. coli* and *S. aureus*, and 30°C for *P. aeruginosa*. The samples underwent a dilution process, and for each dilution, 1 µL of the resulting solution was inoculated. The number of colonies present on MH Petri dishes was then counted after 24 h of incubation. The cell reduction percentage was calculated according to the following formula:

$$\text{Percentage of reduction} = (1 - (C_f/C_i)) \times 100 \text{ in \%} \quad (1)$$

$$C_{i,f} = \log(n/(V*d)) \quad (2)$$

where,

*Ci: Is the initial concentration of microorganisms in contact with NPs before incubation*

*Cf : is the final concentration of microorganisms in contact with NPs after 24 h of incubation*

*n : Is the number of colonies counted*

*V : Is the inoculated volume*

*d : Is the dilution factor*

### 2.6. Kinetic growth of microorganisms

The growth kinetic of microorganisms was monitored using the FLUOstar Omega spectrophotometer at $OD_{600}$. Different concentrations of NPs were aliquoted into a 96-well microplate under the same conditions as before, including the same volume, incubation period, incubation temperature, and bacterial concentration. The $OD_{600}$ measurement was employed as a rapid and cost-effective means to track bacterial growth throughout their culture in liquid media and in contact with NPs. Subsequently, percentage of inhibition, specific growth rate, lag period, and generation time of microorganisms in contact with various ZnO NPs concentrations with diverse sizes were evaluated using $OD_{600}$ data at exponential phase. The formulas used were previously cited in reference [48] and are given below :

$$\mu_x = (\ln(X_2) - \ln(X_1))/(t_2 - t_1) \text{ in } h^{-1} \quad (3)$$

$$G = \ln(2)/\mu_x \text{ in hour} \quad (4)$$

$$\text{Inhibition} = (D_f^{bacteria} - (OD_f^{NPs+bacteria} - OD_{control}^{NPs-bacteria})/OD_f^{bacteria}) \times 100 \text{ in \%} \quad (5)$$

with,

*$\mu_x$ : the specific growth rate*

*$X_2$ and $X_1$ : represent the biomass at the end and the beginning of the exponential growth phase in CFU/mL, respectively*

*$t_2 - t_1$ : the duration of the exponential phase in hour.*

*G : the generation time in hour*

*$OD_f^{bacteria}$ : the final optical density of the positive control with bacteria alone*

*$OD_f^{NPs+bacteria}$ : the final optical density of bacteria with NPs*

*$OD_{control}^{NPs-bacteria}$ : the optical density of NPs alone without bacteria*

The conversion of $OD_{600}$ to CFU/mL was performed using a calibration equation DO = f(X) established within the laboratory for each bacteria under the operational conditions, where X represents the biomass in CFU/mL.

### 2.7. 2D presentation of microorganisms' kinetic growth

The growth kinetics of bacteria in the presence of nanoparticles is represented in 2D using the finite-dimensional Gaussian Process approximation outlined in equation (6), as developed in [50]. Gaussian Processes (GPs) are powerful Bayesian statistical approaches employed as prior distributions over function spaces, used to solve tasks such as regression and classification [51].

$$y_i = \sum_{j_1=1}^{N_1} \ldots \sum_{j_d=1}^{N_d} \xi_{j_1,\ldots,j_d} [\varphi_{j_1}^1(x_i^{(1)}) \times \ldots \times \varphi_{j_d}^d(x_i^{(d)})] \quad (6)$$

where $\xi_{j_1\ldots j_d} = Y(u_{j_1}, \ldots, u_{j_d})$ and $\varphi_{j\zeta}^{\zeta}$ are hat basis functions (see [50] for more details on the properties of these basis functions). With Y a zero-mean GP with covariance function k, i.e., Y~GP (0,k).

### 2.8. Paint formulation with ZnO nanoparticle

To achieve an appropriate dispersion, a preservative-free paint powder was dispersed in sterile distilled water at a mass ratio of 1:1 using a EUROSTAR 20 digital stator rotor emulsifier. Next, a 1w% solution of 22 nm ZnO NPs was added. The paint was then applied to standard glass substrates (VWR) pre-treated with sulfuric acid to ensure better paint adhesion. The paint was left to deeply dry for 24 hours.

### 2.9. Measurement of the antibacterial activity of prepared paints

The antibacterial activity was assessed following the ISO 22196:2011 [52] standard with some modifications. In a typical procedure, strains were cultured on nutrient agar (5.0 g of meat extract, 10.0 g of peptone, 5.0 g of sodium chloride, and 15.0 g of agar powder) and incubated at (35 ± 1) °C for 16 to 24 hours. Subsequently, they were transferred onto fresh slant nutrient agar and further incubated at (35 ± 1) °C for 16 to 20 hours. Bacterial suspensions of 6 Log CFU/mL was then prepared in 1/500 NB nutrient broth (3.0 g of meat extract, 10.0 g of peptone, and 5.0 g of sodium chloride) and a volume of 0.1 mL of the suspension was spread onto paint films. These films were covered with a polypropylene film (2 x 5 cm$^2$) to maintain moisture, and the samples were incubated at 35°C with a relative humidity of ≥ 90% for 24 hours. After incubation, bacterial cells were recovered from the paint films by stirring in 10 mL of Eugon LT broth in presence of 10 g of 1 mm glass beads for 1 minute.

The antibacterial activity was evaluated by colony counting on agar plates. Positive controls were conducted using paint without NPs. The concentration of the bacterial population deposited on the positive control immediately after inoculation was measured to determine the actual concentration deposited. All tests were performed in triplicate, and the antibacterial activity was calculated using the following formula [52] :

$$R = U_t - A_t \quad (7)$$

where,

*R : is the antibacterial activity*

*$U_t$: is the average of the common logarithm of the number of viable bacteria, in cells/cm2, recovered from the untreated test specimens after 24 h*

*$A_t$: is the average of the common logarithm of the number of viable bacteria, in cells/cm2, recovered from the treated test specimens after 24 h.*

### 3. Results
### 3.1. Characterization of ZnO NPs

All ZnO samples are well crystallized in the wurtzite phase, as confirmed by X-ray diffraction (Fig.1, a) and Rietveld refinement (Fig.1, b). The analysis showed stability of the cell parameters (a, c) and reveals varying sizes: 22 nm, 35 nm, and 66 nm (Table 2). Furthermore, transmission electron microscopy (TEM) image provide visual confirmation of the formation of small, well-crystallized ZnO wurtzite NPs (Fig.1, c). Raman spectra further affirm the crystallinity of the samples, exhibiting characteristic Raman modes indicative of the wurtzite structure of ZnO (Fig.1,d).

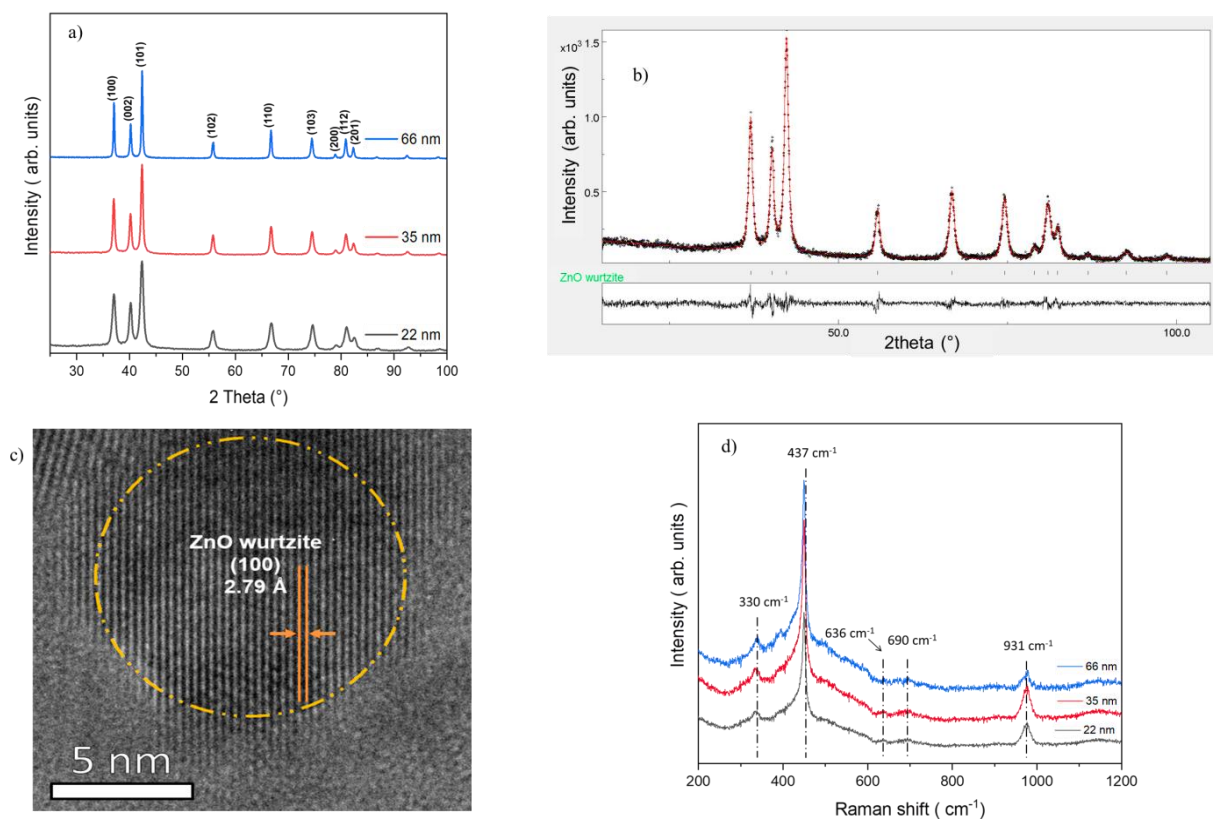

**Figure 1.** a) XRD patterns of samples; b) Rietveld refinement of 22 nm ZnO; c) TEM of 22 nm ZnO dispersed in culture medium and d) Raman spectra of samples.

Table 2 : Crystallographic parameters of the obtained ZnO nanoparticles.

| Crystallite size (nm) | Lattice parameters (Å) | | Microstrain (%) | Sig = GoF |
|---|---|---|---|---|
| | a = b | c | | |
| 22 | 3.25 | 5.20 | 0.002 | 1.18 |
| 35 | 3.25 | 5.21 | 0.001 | 1.21 |
| 66 | 3.25 | 5.21 | 0.001 | 1.60 |

### 3.2. Correlation between ZnO NPs size and minimal inhibitory concentration (MIC)

The MIC represents the lowest concentration of ZnO NPs that inhibits visible bacterial growth following the incubation period. A significant decrease in the MIC of ZnO NPs against *E. coli* was observed as their size decreased, with respective values of 0.45 mg/mL, 0.40 mg/mL, and 0.85 mg/mL for sizes of 22 nm, 35 nm, and 66 nm. Similarly, a reduction in MIC was noted for *P. aeruginosa*, with values decreasing from 1.25 mg/mL, 1.60 mg/mL, to 1.85 mg/mL for sizes of 22 nm, 35 nm, and 66 nm, respectively. The MIC of ZnO NPs against *S.aureus* remained stable at 0.15 mg/mL regardless of their size. These results are presented in Fig. 2.

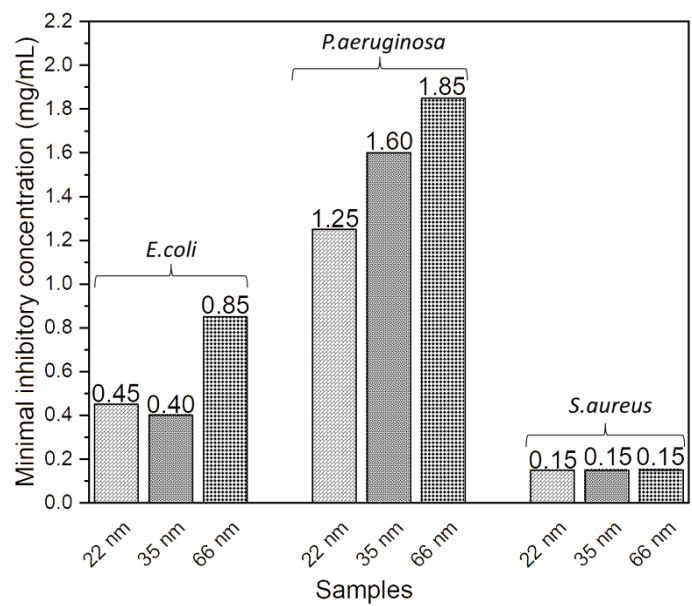

**Figure 2.** Effect of ZnO nanoparticles size in MIC against *E.coli* ; *P.aeruginosa* and *S.aureus.*

### 3.3. Correlation between ZnO NPs size and minimal bactericidal concentration (MBC)

The MBC refers to the lowest concentration of NPs that leads to the death of 99.9% of the initial bacterial population. Its assessment underscores the size-dependent efficacy of ZnO NPs against these bacterial strains (Fig.3). It was observed that the MBC decreases in a size-dependent manner across all three bacteria. Specifically, its values are 0.5 mg/mL, 0.6 mg/mL, and 1.05 mg/mL for sizes of 22 nm, 35 nm, and 66 nm against *E. coli*, respectively. For *P. aeruginosa*, the values are 11 mg/mL, 13 mg/mL, and 15 mg/mL for sizes of 22 nm, 35 nm, and 66 nm, respectively. The same trend was observed against *S. aureus*, with MBC values of 0.30 mg/mL, 1.55 mg/mL, and 1.70 mg/mL for sizes of 22 nm, 35 nm, and 66 nm, respectively.

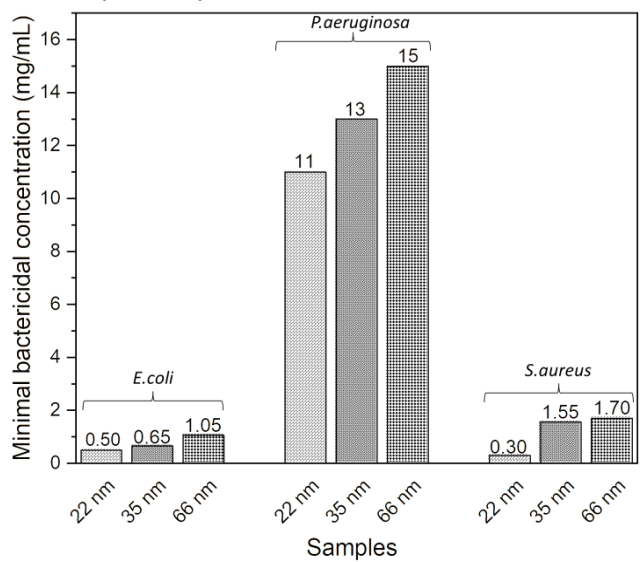

Figure 3. Effect of the size of ZnO NPs on their MBC against *E. coli*, *P. aeruginosa* and *S .aureus*.

### 3.4. ZnO NPs size effect on bacterial growth kinetic

The effects of ZnO NPs with different sizes (22 nm, 35 nm, and 66 nm) and concentrations on bacterial growth inhibition were analyzed. The growth kinetics of bacteria in contact with varying concentrations of NPs of different sizes are visualized in Fig.4, which uses the Gaussian Process (GP) approximation, as outlined in equation (4) and further developed in reference [50]. This GP-based simulation fits well with the observed data, highlighting a rapid reduction in surface optical density as the concentration of NPs increases and the size of NPs decreases. Analysis of $OD_{600}$ growth curves (Fig.4) enabled the determination and comparison of lag phase duration, specific growth rate, and inhibition percentage at different concentrations and sizes.

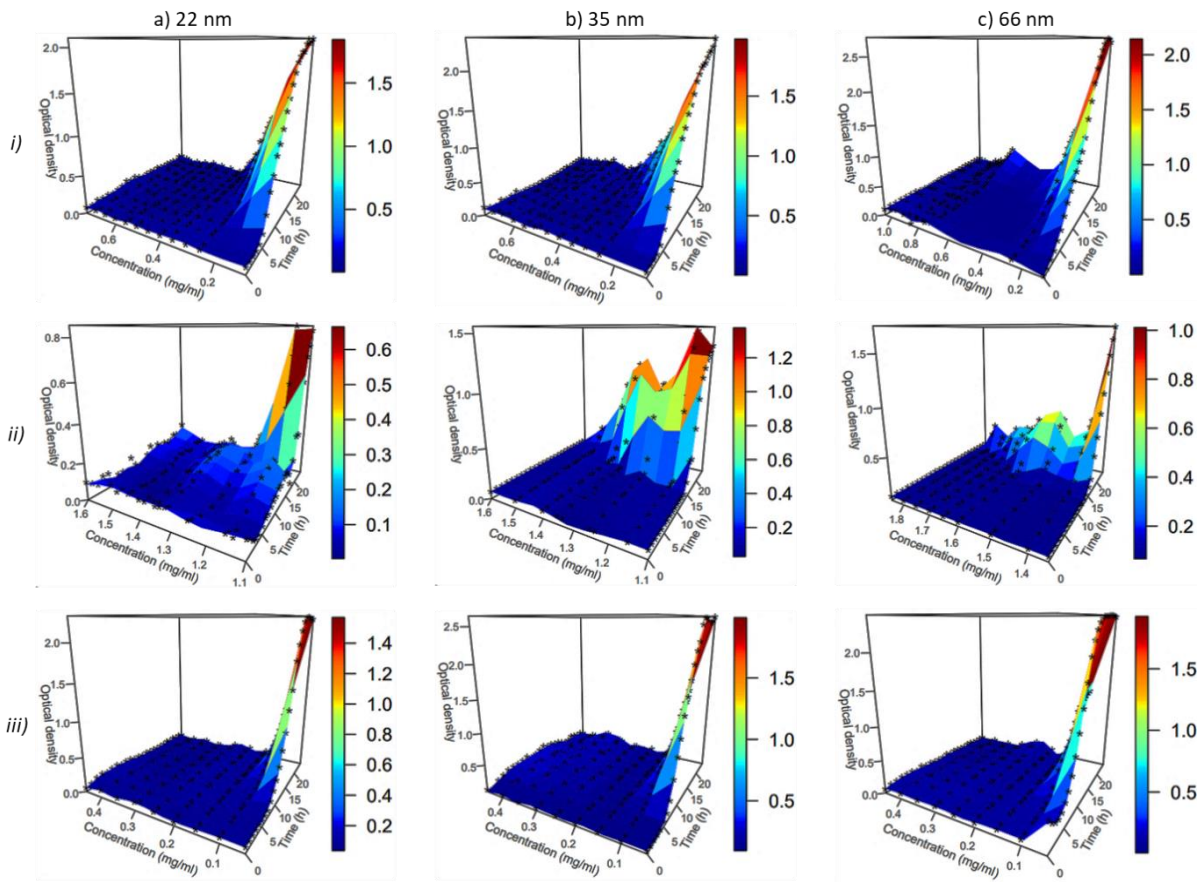

**Figure 4.** $OD_{600}$ growth of i) *E. coli* ; ii) *P. aeruginosa* and iii) *S. aureus* in contact with column a) 22 nm; b) 35 nm and c) 66 nm of ZnO nanoparticles at different concentrations.

The inhibition percentage increased with decreasing NPs size regardless of the bacteria and ZnO concentration (Fig.5, a). While no significant impact is observed with *E. coli* and *S. aureus*, the influence of ZnO NPs size is clearly evident with *P. aeruginosa*. The 22 nm and 35 nm ZnO NPs showed similar inhibition levels on *E. coli* growth across tested concentrations except at 0.3 mg/mL, where inhibition percentages were 74.78%, 76.93%, and 65.59% for 22 nm, 35 nm, and 66 nm NPs, respectively.

The impact of NPs size on inhibition percentage was particularly pronounced in *P. aeruginosa* across all concentrations tested. At 1.35 mg/mL, inhibition percentages decreased from 100%, 59.70% to 39.33% for 22 nm, 35 nm, and 66 nm NPs, respectively. Similarly, at 1.50 mg/mL, inhibition percentages decreased from 100%, 84.72% to 79.72% for the respective NPs sizes as compared with the control. At 1.60 mg/mL, inhibition percentages decreased from 100% for 22 nm and 35 nm NPs to 87.59% for 66 nm NPs.

For *S. aureus*, inhibition percentages were identical for 35 nm and 66 nm NPs across all tested concentrations, at 37% compared to 34% for 22 nm NPs at 0.05 mg/mL. At 0.1 mg/mL, it was 56% compared to 74.10% for 22 nm NPs.

Regarding the specific growth rate, representing bacterial proliferation rate, we observe a general trend across all microorganisms where smaller ZnO decreased the specific growth rate. For *E. coli*, the specific growth rate was 0.67 $h^{-1}$ with 22 nm NPs, compared to 0.8 $h^{-1}$ with 35 nm and 66 nm NPs at 0.1 mg/mL concentration. At 0.25 mg/mL, it decreased to 0.2 $h^{-1}$, 0.3 $h^{-1}$, and 0.49 $h^{-1}$ with 22 nm, 35 nm, and 66 nm NPs, respectively. At 0.3 mg/mL NPs concentration, the specific growth rate was 0.35 $h^{-1}$ with 22 nm and 35 nm NPs, and 0.65 $h^{-1}$ with 66 nm NPs.

*P. aeruginosa* exhibited rates of 0 $h^{-1}$, 0.25 $h^{-1}$, and 0.50 $h^{-1}$ with 22 nm, 35 nm, and 66 nm NPs, respectively, at a concentration of 1.35 mg/mL. At 1.50 mg/mL concentration, the rates were 0 $h^{-1}$, 0.12 $h^{-1}$, and 0.14 $h^{-1}$ with the respective NPs sizes. Finally, at a concentration of 1.60 mg/mL, the specific growth rate of *P. aeruginosa* was 0 $h^{-1}$ for 22 nm and 35 nm NPs, and 0.10 $h^{-1}$ for 66 nm NPs.

For *S. aureus*, at 0.05 mg/mL, the specific growth rate was 0.71 $h^{-1}$, 0.77 $h^{-1}$, and 0.85 $h^{-1}$ with 22 nm, 35 nm, and 66 nm NPs, respectively. On average, it was 0.52 $h^{-1}$ with 22 nm and 35 nm NPs, and 0.67 $h^{-1}$ with 66 nm NPs.

Further elucidation of the impact of NPs size on bacterial growth kinetic was provided by examining the lag phase (Fig.5, c). The lag phase, representing the time for bacteria to initiate growth after inoculation, also exhibits a general trend across all microorganisms, wherein the presence of smaller ZnO particles prolongs the bacterial latency period. At 0.1 mg/mL, the lag phase remained stable across all sizes for *E. coli*, averaging 4 hours. However, notable reductions were observed at higher concentrations. For instance, at 0.25 mg/mL, the lag phase of *E. coli* decreased from 8.5 hours, 10 hours to 4 hours with 22 nm, 35 nm, and 66 nm NPs, respectively. Similarly, at 0.30 mg/mL, it further decreased to 12 hours, 14.5 hours, and 4.5 hours with the respective NPs sizes.

In the case of *P. aeruginosa*, at a concentration of 1.35 mg/mL, the lag phase reduced from 23 hours to 13 hours with 22 nm, 35 nm, and 66 nm NPs. This reduction was also observed at 1.50 mg/mL, where the lag phase was 24 hours, 21 hours, and 15 hours with 22 nm, 35 nm, and 66 nm NPs, respectively. Similarly, at 1.60 mg/mL NPs concentration, the lag phase remained 24 hours for both 22 nm and 35 nm sizes, and decreased to 18 hours for 66 nm NPs.

Similarly, for *S. aureus*, distinct differences in lag phase duration were observed at different concentrations and NPs sizes. At a concentration of 0.05 mg/mL, the lag phase persisted for 9 hours with 22 nm NPs, while it was reduced to 7 hours with 35 nm and 66 nm NPs. Conversely, at a concentration of 0.1 mg/mL, the lag phase extended to 15 hours with 22 nm NPs, while it remained at 10 hours with 35 nm and 66 nm NPs.

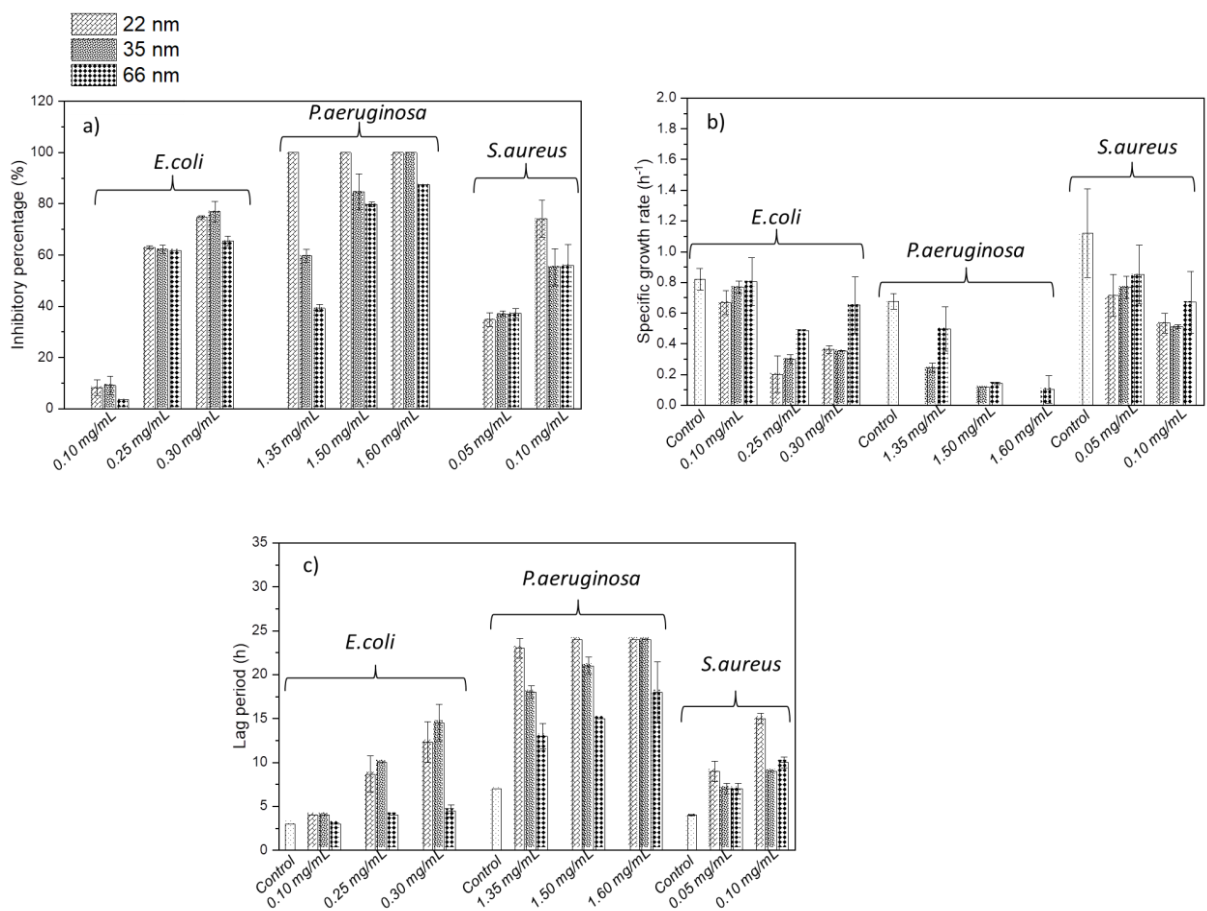

**Figure 5.** ZnO NPs size and concentration effect on the growth parameter of *E. coli*; *P. aeruginosa* and *S. aureus* : a) Inhibitory percentage; b) Specific growth rate ; and c) Lag period.

**Figure 6** shows an increase in the bacterial growth inhibition with decreasing NPs size for all microorganisms, at different ZnO nanoparticle doses. For *E. coli*, inhibition was consistent with 22 nm and 35 nm nanoparticles, and higher than with 66 nm nanoparticles. *P. aeruginosa* inhibition increases with decreasing size. Furthermore, as nanoparticle concentration increases, inhibition tends to converge for all three nanoparticle sizes. As for *S. aureus*, inhibition is consistent with 66 nm and 35 nm nanoparticles, but weaker in relation to contact with 22 nm nanoparticles. These results confirm those obtained for growth parameters, MIC and MBC.

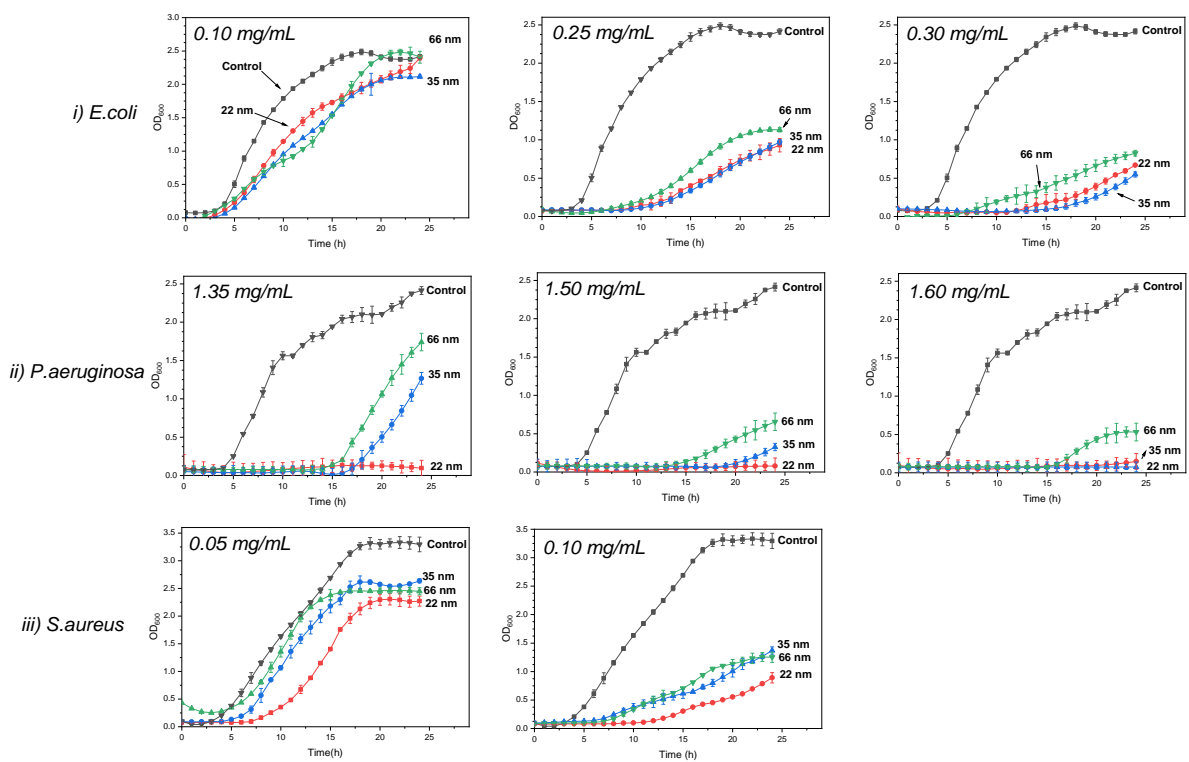

**Figure 6.** Effect of ZnO NPs size on bacterial growth kinetics.

### 3.5. Antibacterial activity of prepared paint

The paint formulation incorporating 22 nm ZnO demonstrates superior efficacy against *S. aureus* and *E. coli* in contrast to *P. aeruginosa* (Fig.7). Remarkably, we have the same trend of results compared to those of the nanoparticle suspension; *P.aeruginosa* is the least sensitive to NPs. It is noteworthy that the detachment of paint from the slides occurred subsequent to the vortexing step used to remove bacteria from the glass slide.

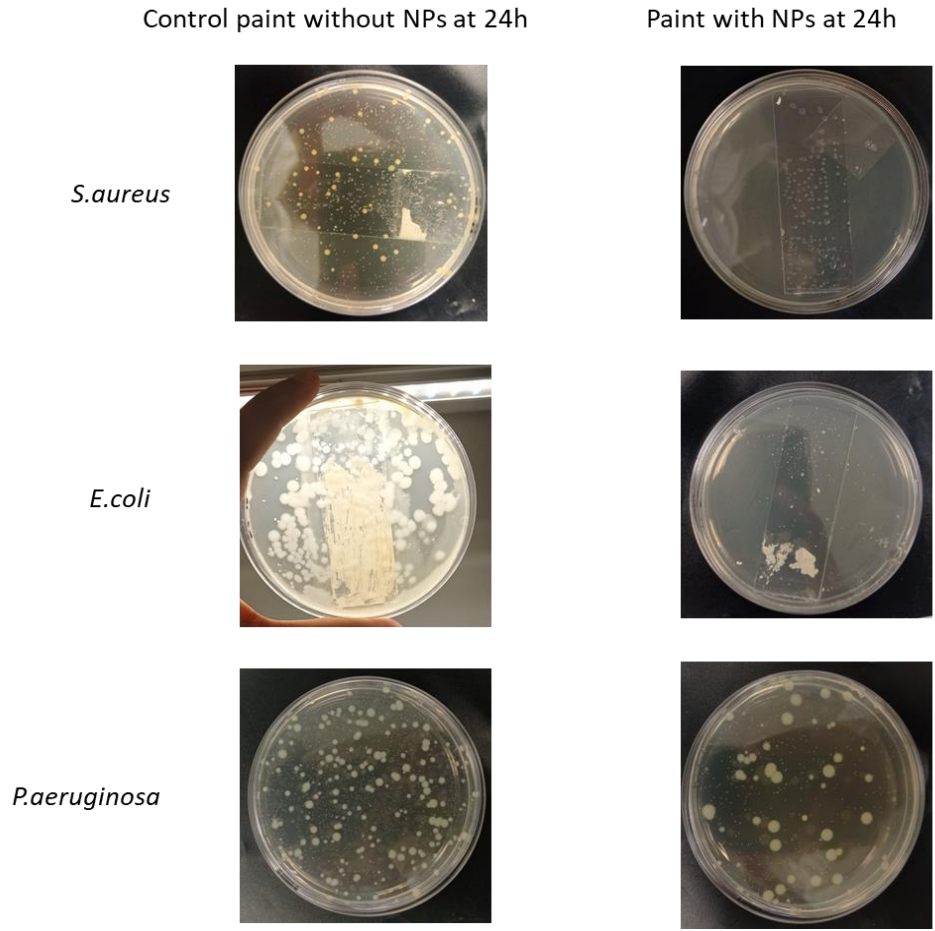

**Figure 7.** Evaluation of antibacterial properties of paint containing ZnO nanoparticles against *S.aureus* ; *E.coli* and *P.aeruginosa*.

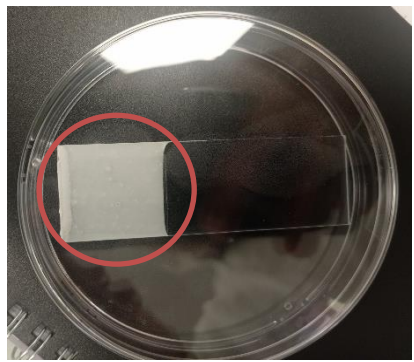

**Figure 8.** An illustration of a paint film with 22 nm nanoparticles, onto which 0.1 mL of bacterial suspension is spread, covered with a polypropylene film.

## 4. Discussion

The most important findings of this study can be summarized as follows: Firstly, decreasing the size of ZnO NPs (22 nm, 35 nm, and 66 nm) enhanced their antibacterial efficacy by lowering their minimum inhibitory concentration (MIC) and minimum bactericidal concentration (MBC). Notably, *P. aeruginosa* exhibited a lower susceptibility to NPs compared to the other bacterial strains, necessitating higher concentrations for bactericidal effect. Furthermore, the reduction in MIC and MBC with decreasing NPs size was particularly pronounced when dealing with *P. aeruginosa*.

Several investigations have explored the effects of ZnO NPs on MIC and MBC values [9,10,36,37,53,54,54]. Consistent with our findings, most researchers have observed a significant enhancement in antibacterial effectiveness with smaller NPs. For instance, Pasquet et al. [9] reported superior antimicrobial activity with smaller ZnO crystals against *E. coli* and *P. aeruginosa*, noting a proportional increase in MIC and MBC with larger NPs.

Álvarez-Chimal et al. [55] have observed consistent findings, indicating that ZnO NPs sized at 7 nm, 10 nm, 16 nm, and 30 nm show increased MBC against both *S. aureus* and *E. coli* as their size increases. Remarkably, MBC values reported in their study, ranging from 10 to 80 mg/mL, are higher than those observed in the present investigation, likely due to differences in NP composition. However, our results regarding *S. aureus* diverge from those of Lallo Da Silva et al. [43] and Nicole et al. [10], who have documented elevated MIC values with larger NPs against *S.aureus* strains. Additionally, Palanikumar et al. [36] reported that ZnO NPs with sizes of 15 nm, 25 nm, and 38 nm exhibited identical MICs against *S. aureus* MRSA, with a value of 0.2 mg/mL, aligning with our findings. Despite *S. aureus* consistently displaying similar MIC values regardless of NPs size, we observed a decline in cell viability, as indicated by MBC measurements, with decreasing particle size. The lack of variation in MIC of NPs against *S. aureus* based on their size may be attributed to challenges in turbidity detection arising from the small size of bacterial cells.

Previous studies have suggested that *E. coli* is more susceptible to ZnO NPs than *P. aeruginosa*, while *S. aureus* is more sensitive than *P. aeruginosa*. The variation in susceptibility among these bacteria can be attributed to a complex interplay of factors, including the production of extracellular polymeric substances (EPS) [56,57], detoxification systems, and specific metabolic and genetic responses. The extensive coding capacity of the *P. aeruginosa* genome enables remarkable metabolic adaptability and versatility in response to environmental changes [58,59].

Secondly, the study suggested that the diminution in the size of ZnO NPs is able to slow down the bacterial growth. Few studies in the literature have delved into the role of ZnO NPs size in the kinetic growth of bacteria. Nevertheless, Baek et al. discovered that the growth inhibition rate (%) of *E. coli* increased as the size of ZnO NPs decreased (12.7 nm, 15.7 nm, and 17.2 nm)[60]. These results are in line with findings from Nicole et al. [10] and Raghupathi et al. [35], who have similarly observed an increase in growth inhibition against *S.aureus* and *E.coli* with smaller ZnO NPs sizes.

The antibacterial activity of ZnO NPs is attributed to several mechanisms, including the release of reactive oxygen species (ROS), destruction of the cell membrane, and internalization of NPs into bacterial cells. The reduction in NPs size can influence these mechanisms and slow bacterial growth in several ways. The heightened reactivity of smaller-sized NPs due to their larger specific surface area can promote increased ROS release upon interaction with bacteria, leading to greater oxidation and deterioration of essential cellular components, thus slowing bacterial growth. Also, the reduced size of NPs can result in more efficient internalization into bacterial cells. Lastly, the decrease in NPs size can also increase the likelihood of direct contact with the cell membrane, potentially causing more significant membrane disruption and leakage of cellular components, thereby compromising bacterial viability. Furthermore, the similar antibacterial activity of the 35 nm and 66 nm NPs against *S. aureus*, in contrast to the distinct activity observed with the smaller 22 nm NPs, which demonstrate greater efficacy specifically against *S. aureus*, may be attributed to the unique characteristic of *S. aureus* to form aggregates in suspension[61]. There appears to be a maximum size beyond which NPs cannot penetrate between these aggregates, thus maintaining unchanged antibacterial activity. Furthermore, the inhibitory effects of NPs on *P. aeruginosa* growth tend to converge as their concentration increases, probably because the minimum inhibitory concentration (MIC) is close.

Thirdly, ZnO nano-based paints exhibit bactericidal activity against *E. coli* and *S. aureus*. These findings align with literature demonstrating the effectiveness of paints containing ZnO nanoparticles against various bacteria, including *E. coli* and *S. aureus* [62–64]. For example, Fiori et al. [62] found that the most significant antimicrobial effect against *S. aureus*, assessed using an agar diffusion test, was achieved with the paint formulated with 1.2% of 9 nm nanoZnO.

## 5. Conclusion

The reduction in the size of ZnO nanoparticles enhances their antibacterial activity against *E. coli*, *P. aeruginosa*, and *S. aureus*. Additionally, it decelerates the growth kinetics of bacteria by reducing their specific growth rate, prolonging their lag time, and increasing their inhibition percentage. This study emphasizes the critical importance of ZnO NPs size in their ability to inhibit and eradicate pathogenic Gram-negative and Gram-positive bacteria. Furthermore, the application of these ZnO nanoparticles in paint formulations presents a promising strategy to combat nosocomial infections. A prospective direction for this study is to investigate the safety of our ZnO NPs and the antibacterial activity of thin films of ZnO nanoparticles.

**Data availability**

Data will be available on request.

**Acknowledgements**

The authors would like to acknowledge Paule Arielle and Amelia Murray from the School Ecole de Biologie Industrielle (EBI) for assistance in the determination of MBC by counting colonies, and Maria Konstantakoupoulou from the Laboratoire des Sciences des Procédés et des Matériaux (LSPM) laboratory for help in obtaining the MET image. The authors would also like to acknowledge Prof. Andrei Kanaev of the Galilée institute for reviewing this article.

**Fundings**

This work was supported by the CNRS-LSPM UPR 3407, University Sorbonne Paris Nord and the Ecole de Biologie Industrielle (EBI).

**References**

[1] A.S. Baviskar, K.I. Khatib, D. Rajpal, H.C. Dongare, Nosocomial infections in surgical intensive care unit: A retrospective single-center study, Int J Crit Illn Inj Sci 9 (2019) 16–20. https://doi.org/10.4103/IJCIIS.IJCIIS_57_18.

[2] A. Pugazhendhi, S. Vasantharaj, S. Sathiyavimal, R.K. Raja, I. Karuppusamy, M. Narayanan, S. Kandasamy, K. Brindhadevi, Organic and inorganic nanomaterial coatings for the prevention of microbial growth and infections on biotic and abiotic surfaces, Surface and Coatings Technology 425 (2021) 127739. https://doi.org/10.1016/j.surfcoat.2021.127739.

[3] N. Beyth, Y. Houri-Haddad, A. Domb, W. Khan, R. Hazan, Alternative antimicrobial approach: nano-antimicrobial materials, Evid Based Complement Alternat Med 2015 (2015) 246012. https://doi.org/10.1155/2015/246012.

[4] L. Palanikumar, S.N. Ramasamy, C. Balachandran, Size-dependent antimicrobial response of zinc oxide nanoparticles, IET Nanobiotechnol. 8 (2014) 111–117. https://doi.org/10.1049/iet-nbt.2012.0008.

[5] J. Sawai, Quantitative evaluation of antibacterial activities of metallic oxide powders (ZnO, MgO and CaO) by conductimetric assay, Journal of Microbiological Methods 54 (2003) 177–182. https://doi.org/10.1016/S0167-7012(03)00037-X.

[6] S.K. Arya, S. Saha, J.E. Ramirez-Vick, V. Gupta, S. Bhansali, S.P. Singh, Recent advances in ZnO nanostructures and thin films for biosensor applications: Review, Analytica Chimica Acta 737 (2012) 1–21. https://doi.org/10.1016/j.aca.2012.05.048.

[7] S. Jiang, K. Lin, M. Cai, ZnO Nanomaterials: Current Advancements in Antibacterial Mechanisms and Applications, Front. Chem. 8 (2020). https://doi.org/10.3389/fchem.2020.00580.

[8] M. Izzi, M.C. Sportelli, L. Torsi, R.A. Picca, N. Cioffi, Synthesis and Antimicrobial Applications of ZnO Nanostructures: A Review, ACS Appl. Nano Mater. 6 (2023) 10881–10902. https://doi.org/10.1021/acsanm.3c01432.

[9] J. Pasquet, Y. Chevalier, E. Couval, D. Bouvier, G. Noizet, C. Morlière, M.-A. Bolzinger, Antimicrobial activity of zinc oxide particles on five micro-organisms of the Challenge Tests related to their physicochemical properties, Int J Pharm 460 (2014) 92–100. https://doi.org/10.1016/j.ijpharm.2013.10.031.

[10] N. Jones, B. Ray, K.T. Ranjit, A.C. Manna, Antibacterial activity of ZnO nanoparticle suspensions on a broad spectrum of microorganisms, FEMS Microbiology Letters 279 (2008) 71–76. https://doi.org/10.1111/j.1574-6968.2007.01012.x.

[11] A.C. Janaki, E. Sailatha, S. Gunasekaran, Synthesis, characteristics and antimicrobial activity of ZnO nanoparticles, Spectrochimica Acta Part A: Molecular and Biomolecular Spectroscopy 144 (2015) 17–22. https://doi.org/10.1016/j.saa.2015.02.041.

[12] R.K. Dutta, B.P. Nenavathu, M.K. Gangishetty, A.V.R. Reddy, Studies on antibacterial activity of ZnO nanoparticles by ROS induced lipid peroxidation, Colloids and Surfaces B: Biointerfaces 94 (2012) 143–150. https://doi.org/10.1016/j.colsurfb.2012.01.046.

[13] A. Abdal Dayem, M.K. Hossain, S.B. Lee, K. Kim, S.K. Saha, G.-M. Yang, H.Y. Choi, S.-G. Cho, The Role of Reactive Oxygen Species (ROS) in the Biological Activities of Metallic Nanoparticles, International Journal of Molecular Sciences 18 (2017) 120. https://doi.org/10.3390/ijms18010120.

[14] A. Lipovsky, Z. Tzitrinovich, H. Friedmann, G. Applerot, A. Gedanken, R. Lubart, EPR Study of Visible Light-Induced ROS Generation by Nanoparticles of ZnO, J. Phys. Chem. C 113 (2009) 15997–16001. https://doi.org/10.1021/jp904864g.

[15] N. Babayevska, Ł. Przysiecka, I. Iatsunskyi, G. Nowaczyk, M. Jarek, E. Janiszewska, S. Jurga, ZnO size and shape effect on antibacterial activity and cytotoxicity profile, Sci Rep 12 (2022) 8148. https://doi.org/10.1038/s41598-022-12134-3.

[16] S. Magder, Reactive oxygen species: toxic molecules or spark of life?, Crit Care 10 (2006) 208. https://doi.org/10.1186/cc3992.

[17] B. Abebe, E.A. Zereffa, A. Tadesse, H.C.A. Murthy, A Review on Enhancing the Antibacterial Activity of ZnO: Mechanisms and Microscopic Investigation, Nanoscale Res Lett 15 (2020) 190. https://doi.org/10.1186/s11671-020-03418-6.

[18] U. Manzoor, S. Siddique, R. Ahmed, Z. Noreen, H. Bokhari, I. Ahmad, Antibacterial, Structural and Optical Characterization of Mechano-Chemically Prepared ZnO Nanoparticles, PLoS ONE 11 (2016) e0154704. https://doi.org/10.1371/journal.pone.0154704.

[19] S. Raha, Md. Ahmaruzzaman, ZnO nanostructured materials and their potential applications: progress, challenges and perspectives, Nanoscale Adv. 4 (2022) 1868–1925. https://doi.org/10.1039/D1NA00880C.

[20] Y. Liu, L. He, A. Mustapha, H. Li, Z.Q. Hu, M. Lin, Antibacterial activities of zinc oxide nanoparticles against Escherichia coli O157:H7, J Appl Microbiol 107 (2009) 1193–1201. https://doi.org/10.1111/j.1365-2672.2009.04303.x.

[21] C.R. Mendes, G. Dilarri, C.F. Forsan, V. de M.R. Sapata, P.R.M. Lopes, P.B. de Moraes, R.N. Montagnolli, H. Ferreira, E.D. Bidoia, Antibacterial action and target mechanisms of zinc oxide nanoparticles against bacterial pathogens, Sci Rep 12 (2022) 2658. https://doi.org/10.1038/s41598-022-06657-y.

[22] Happy Agarwal, Soumya Menon, S. Venkat Kumar, S. Rajeshkumar, Mechanistic study on antibacterial action of zinc oxide nanoparticles synthesized using green route, Chemico-Biological Interactions 286 (2018) 60–70. https://doi.org/10.1016/j.cbi.2018.03.008.

[23] S. Thakur, S. Neogi, Effect of doped ZnO nanoparticles on bacterial cell morphology and biochemical composition, Appl Nanosci 11 (2021) 159–171. https://doi.org/10.1007/s13204-020-01592-8.

[24] Z. Xin, Q. He, S. Wang, X. Han, Z. Fu, X. Xu, X. Zhao, Recent Progress in ZnO-Based Nanostructures for Photocatalytic Antimicrobial in Water Treatment: A Review, Applied Sciences 12 (2022) 7910. https://doi.org/10.3390/app12157910.

[25] P.K. Stoimenov, R.L. Klinger, G.L. Marchin, K.J. Klabunde, Metal Oxide Nanoparticles as Bactericidal Agents, Langmuir 18 (2002) 6679–6686. https://doi.org/10.1021/la0202374.

[26] P. Bhattacharya, A. Dey, S. Neogi, An insight into the mechanism of antibacterial activity by magnesium oxide nanoparticles, J. Mater. Chem. B 9 (2021) 5329–5339. https://doi.org/10.1039/D1TB00875G.

[27] null Happy Agarwal, null Soumya Menon, S. Venkat Kumar, S. Rajeshkumar, Mechanistic study on antibacterial action of zinc oxide nanoparticles synthesized using green route, Chem Biol Interact 286 (2018) 60–70. https://doi.org/10.1016/j.cbi.2018.03.008.

[28] S. Sharma, K. Kumar, N. Thakur, S. Chauhan, M.S. Chauhan, The effect of shape and size of ZnO nanoparticles on their antimicrobial and photocatalytic activities: a green approach, Bull Mater Sci 43 (2019) 20. https://doi.org/10.1007/s12034-019-1986-y.

[29] F. Shahmohammadi Jebel, H. Almasi, Morphological, physical, antimicrobial and release properties of ZnO nanoparticles-loaded bacterial cellulose films, Carbohydrate Polymers 149 (2016) 8–19. https://doi.org/10.1016/j.carbpol.2016.04.089.

[30] Q. Cai, Y. Gao, T. Gao, S. Lan, O. Simalou, X. Zhou, Y. Zhang, C. Harnoode, G. Gao, A. Dong, Insight into Biological Effects of Zinc Oxide Nanoflowers on Bacteria: Why Morphology Matters, ACS Appl. Mater. Interfaces 8 (2016) 10109–10120. https://doi.org/10.1021/acsami.5b11573.

[31] N. Talebian, S.M. Amininezhad, M. Doudi, Controllable synthesis of ZnO nanoparticles and their morphology-dependent antibacterial and optical properties, Journal of Photochemistry and Photobiology B: Biology 120 (2013) 66–73. https://doi.org/10.1016/j.jphotobiol.2013.01.004.

[32] P.B. Jaiswal, S. Jejurikar, A. Mondal, B. Pushkar, S. Mazumdar, Antibacterial Effects of ZnO Nanodisks: Shape Effect of the Nanostructure on the Lethality in Escherichia coli, Appl Biochem Biotechnol 195 (2023) 3067–3095. https://doi.org/10.1007/s12010-022-04265-0.

[33] L. Motelica, O.-C. Oprea, B.-S. Vasile, A. Ficai, D. Ficai, E. Andronescu, A.M. Holban, Antibacterial Activity of Solvothermal Obtained ZnO Nanoparticles with Different Morphology and Photocatalytic Activity against a Dye Mixture: Methylene Blue, Rhodamine B and Methyl Orange, International Journal of Molecular Sciences 24 (2023) 5677. https://doi.org/10.3390/ijms24065677.

[34] B. Lallo da Silva, B.L. Caetano, B.G. Chiari-Andréo, R.C.L.R. Pietro, L.A. Chiavacci, Increased antibacterial activity of ZnO nanoparticles: Influence of size and surface modification, Colloids and Surfaces B: Biointerfaces 177 (2019) 440–447. https://doi.org/10.1016/j.colsurfb.2019.02.013.

[35] K.R. Raghupathi, R.T. Koodali, A.C. Manna, Size-Dependent Bacterial Growth Inhibition and Mechanism of Antibacterial Activity of Zinc Oxide Nanoparticles, Langmuir 27 (2011) 4020–4028. https://doi.org/10.1021/la104825u.

[36] L. Palanikumar, S.N. Ramasamy, C. Balachandran, Size-dependent antimicrobial response of zinc oxide nanoparticles, IET Nanobiotechnol 8 (2014) 111–117. https://doi.org/10.1049/iet-nbt.2012.0008.

[37] A. Azam, A.S. Ahmed, M. Oves, M. Khan, A. Memic, Size-dependent antimicrobial properties of CuO nanoparticles against Gram-positive and -negative bacterial strains, International Journal of Nanomedicine 7 (2012) 3527–3535. https://doi.org/10.2147/IJN.S29020.

[38] A. Awasthi, P. Sharma, L. Jangir, Kamakshi, G. Awasthi, K.K. Awasthi, K. Awasthi, Dose dependent enhanced antibacterial effects and reduced biofilm activity against *Bacillus subtilis* in presence of ZnO nanoparticles, Materials Science and Engineering: C 113 (2020) 111021. https://doi.org/10.1016/j.msec.2020.111021.

[39] R. Dadi, E. Kerignard, M. Traoré, C. Mielcareck, A. Kanaev, R. Azouani, Evaluation of Antibacterial Efficiency of Zinc Oxide Thin Films Nanoparticles against Nosocomial Bacterial Strains, Chemical Engineering Transactions (2021). https://doi.org/10.3303/CET2184003.

[40] L.C. Ann, S. Mahmud, S.K.M. Bakhori, A. Sirelkhatim, D. Mohamad, H. Hasan, A. Seeni, R.A. Rahman, Effect of surface modification and UVA photoactivation on antibacterial bioactivity of zinc oxide powder, Applied Surface Science 292 (2014) 405–412. https://doi.org/10.1016/j.apsusc.2013.11.152.

[41] F. Wang, J. Qi, L. Zhu, Ag/MoS2 nanozyme-modified ZnO nanopillar surface for enhanced synergistic mechanical and chemical antibacterial activity, Colloids and Surfaces A: Physicochemical and Engineering Aspects 687 (2024) 133494. https://doi.org/10.1016/j.colsurfa.2024.133494.

[42] Y. Zaman, M.Z. Ishaque, K. Waris, M. Shahzad, A.B. Siddique, M.I. Arshad, H. Zaman, H.M. Ali, F. Kanwal, M. Aslam, M. Mustaqeem, Modified physical properties of Ni doped ZnO NPs as potential photocatalyst and antibacterial agents, Arabian Journal of Chemistry 16 (2023) 105230. https://doi.org/10.1016/j.arabjc.2023.105230.

[43] B. Lallo da Silva, B.L. Caetano, B.G. Chiari-Andréo, R.C.L.R. Pietro, L.A. Chiavacci, Increased antibacterial activity of ZnO nanoparticles: Influence of size and surface modification, Colloids and Surfaces B: Biointerfaces 177 (2019) 440–447. https://doi.org/10.1016/j.colsurfb.2019.02.013.

[44] L. Xu, H.-W. Liang, Y. Yang, S.-H. Yu, Stability and Reactivity: Positive and Negative Aspects for Nanoparticle Processing, Chem. Rev. 118 (2018) 3209–3250. https://doi.org/10.1021/acs.chemrev.7b00208.

[45] J. Amodeo, L. Pizzagalli, Modeling the mechanical properties of nanoparticles: a review, Comptes Rendus. Physique 22 (2021) 35–66. https://doi.org/10.5802/crphys.70.

[46] A. Sukhanova, S. Bozrova, P. Sokolov, M. Berestovoy, A. Karaulov, I. Nabiev, Dependence of Nanoparticle Toxicity on Their Physical and Chemical Properties, Nanoscale Res Lett 13 (2018) 44. https://doi.org/10.1186/s11671-018-2457-x.

[47] S. Nair, A. Sasidharan, V.V. Divya Rani, D. Menon, S. Nair, K. Manzoor, S. Raina, Role of size scale of ZnO nanoparticles and microparticles on toxicity toward bacteria and osteoblast cancer cells, J Mater Sci: Mater Med 20 (2009) 235–241. https://doi.org/10.1007/s10856-008-3548-5.

[48] I. El-Habib, A. Roynette, H. Morakchi-Goudjil, A. Lemarchand, M. Christine, R. Azouani, M. Traore, Synthesis by soft chemistry of size-controlled zinc oxide (ZnO) nanocrystals for antimicrobial applications, MATEC Web of Conferences 379 (2023). https://doi.org/10.1051/matecconf/202337906003.

[49] Z. Luo, M. Zhu, M. Guo, Z. Lian, W. Tong, J. Wang, B. Zhang, W. Wei, Ultrasonic-Assisted Dispersion of ZnO Nanoparticles and Its Inhibition Activity to Trichoderma viride, J Nanosci Nanotechnol 18 (2018) 2352–2360. https://doi.org/10.1166/jnn.2018.14397.

[50] H. Maatouk, X. Bay, Gaussian Process Emulators for Computer Experiments with Inequality Constraints, Math Geosci 49 (2017) 557–582. https://doi.org/10.1007/s11004-017-9673-2.

[51] C.E. Rasmussen, C.K.I. Williams, Gaussian Processes for Machine Learning, The MIT Press, 2005. https://doi.org/10.7551/mitpress/3206.001.0001.

[52] ISO 22196:2011(en), Measurement of antibacterial activity on plastics and other non-porous surfaces, (n.d.). https://www.iso.org/obp/ui/#iso:std:iso:22196:ed-2:v1:en (accessed March 18, 2024).

[53] B. Lallo da Silva, M.P. Abuçafy, E. Berbel Manaia, J.A. Oshiro Junior, B.G. Chiari-Andréo, R.C.R. Pietro, L.A. Chiavacci, Relationship Between Structure And Antimicrobial Activity Of Zinc Oxide Nanoparticles: An Overview, International Journal of Nanomedicine 14 (2019) 9395–9410. https://doi.org/10.2147/IJN.S216204.

[54] G. Applerot, A. Lipovsky, R. Dror, N. Perkas, Y. Nitzan, R. Lubart, A. Gedanken, Enhanced Antibacterial Activity of Nanocrystalline ZnO Due to Increased ROS-Mediated Cell Injury, Advanced Functional Materials 19 (2009) 842–852. https://doi.org/10.1002/adfm.200801081.

[55] R. Álvarez-Chimal, V.I. García-Pérez, M.A. Álvarez-Pérez, R. Tavera-Hernández, L. Reyes-Carmona, M. Martínez-Hernández, J.Á. Arenas-Alatorre, Influence of the particle size on the antibacterial activity of green synthesized zinc oxide nanoparticles using *Dysphania ambrosioides* extract, supported by molecular docking analysis, Arabian Journal of Chemistry 15 (2022) 103804. https://doi.org/10.1016/j.arabjc.2022.103804.

[56] . N., R. Durgesh, S. Sakthivel, Evaluation of extra-polymeric substance (EPS) of Pseudomonas aeruginosa and Escherica coli in water distribution system, Asain Journal of Microbial Biotec and Environmental Science 12 (2010) 199–204.

[57] R. Grossich, M. Lemos Vilches, C.S. Costa, M. Pezzoni, Role of Pel and Psl polysaccharides in the response of Pseudomonas aeruginosa to environmental challenges: oxidative stress agents (UVA, H2O2, sodium hypochlorite) and its competitor Staphylococcus aureus, Microbiology 169 (2023) 001301. https://doi.org/10.1099/mic.0.001301.

[58] Z. Pang, R. Raudonis, B.R. Glick, T.-J. Lin, Z. Cheng, Antibiotic resistance in Pseudomonas aeruginosa: mechanisms and alternative therapeutic strategies, Biotechnology Advances 37 (2019) 177–192. https://doi.org/10.1016/j.biotechadv.2018.11.013.

[59] J. Klockgether, N. Cramer, L. Wiehlmann, C. Davenport, B. Tümmler, Pseudomonas aeruginosa Genomic Structure and Diversity, Frontiers in Microbiology 2 (2011). https://www.frontiersin.org/articles/10.3389/fmicb.2011.00150 (accessed October 1, 2023).

[60] S. Baek, S.H. Joo, N. Kumar, M. Toborek, Antibacterial effect and toxicity pathways of industrial and sunscreen ZnO nanoparticles on *Escherichia coli*, Journal of Environmental Chemical Engineering 5 (2017) 3024–3032. https://doi.org/10.1016/j.jece.2017.06.009.

[61] J. Haaber, M.T. Cohn, D. Frees, T.J. Andersen, H. Ingmer, Planktonic Aggregates of Staphylococcus aureus Protect against Common Antibiotics, PLOS ONE 7 (2012) e41075. https://doi.org/10.1371/journal.pone.0041075.

[62] J. Fiori, L. Silva, K.C. Picolli, R. Ternus, J. Ilha, F. Decalton, J.M. de Mello, H. Riella, M. Fiori, Zinc Oxide Nanoparticles as Antimicrobial Additive for Acrylic Paint, Materials Science Forum 899 (2017) 248–253. https://doi.org/10.4028/www.scientific.net/MSF.899.248.

[63] W. Agustin, M.T. Albari, M.A. Ghifari, M.R. Ghifari, D. Purnamasari, R.S. Mandeli, The antibacterial properties of paint with the addition of ZnO nanoparticles, AIP Conference Proceedings 3001 (2024) 030004. https://doi.org/10.1063/5.0184163.

[64] H. Foudi, A. Soukeur, G. Rekhila, M. Trari, M. Amara, Synthesis and characterization of ZnO nanoparticles for antibacterial paints, Chem. Pap. 77 (2023) 1489–1496. https://doi.org/10.1007/s11696-022-02565-7.